\documentclass{elsart}
\usepackage{natbib}
\usepackage{graphicx}
\usepackage{amssymb}

\begin{document}
\bibliographystyle{unsrt}

\begin{frontmatter}

\title{Development of a Time Projection Chamber Using Gas Electron Multipliers (GEM-TPC)}

\author[CNS]{S.X.~Oda\corauthref{cor1}},
\corauth[cor1]{Corresponding author.}
\ead{oda@cns.s.u-tokyo.ac.jp}
\author[CNS]{H.~Hamagaki},
\author[CNS]{K.~Ozawa},
\author[CNS]{M.~Inuzuka\thanksref{NRICPT}},
\author[CNS]{T.~Sakaguchi\thanksref{BNL}},
\author[CNS]{T.~Isobe},
\author[CNS]{T.~Gunji},
\author[CNS]{Y.~Morino},
\author[CNS]{S.~Saito},
\author[Waseda]{Y.L.~Yamaguchi\thanksref{CNSt}},
\author[KEK]{S.~Sawada}, 
\author[RIKEN]{S.~Yokkaichi}

\address[CNS]{Center for Nuclear Study, Graduate School of Science, University of Tokyo, 7-3-1 Hongo, Bunkyo, Tokyo 113-0033, Japan }
\address[Waseda]{Waseda University, Advanced Research Institute for Science and Engineering, 17 Kikui-cho, Shinjuku-ku, Tokyo 162-0044, Japan }
\address[KEK]{KEK, High Energy Accelerator Research Organization, Tsukuba-shi, Ibaraki-ken 305-0801, Japan }
\address[RIKEN]{RIKEN (The Institute of Physical and Chemical Research), Wako, Saitama 351-0198, Japan }

\thanks[NRICPT]{Now at National Research Institute for Cultural Properties, Tokyo, 13-43 Ueno Park, Taito-ku, Tokyo 110-8713, Japan. }
\thanks[BNL]{Now at Brookhaven National Laboratory, Upton, NY 11973-5000, U.S. }
\thanks[CNSt]{Now at Center for Nuclear Study, Graduate School of Science, University of Tokyo, 7-3-1 Hongo, Bunkyo, Tokyo 113-0033, Japan. }

\begin{abstract}
We developed a prototype time projection chamber using gas electron multipliers (GEM-TPC) for high energy heavy ion collision experiments. 
To investigate its performance, we conducted a beam test with 3 kinds of gases (Ar(90\%)-CH$_4$(10\%), Ar(70\%)-C$_2$H$_6$(30\%) and CF$_4$). 
Detection efficiency of 99\%, and spatial resolution of 79~$\mu$m in the pad-row direction and 313~$\mu$m in the drift direction were achieved.
The test results show that the GEM-TPC meets the requirements for high energy heavy ion collision experiments. 
The configuration and performance of the GEM-TPC are described. 
\end{abstract}

\begin{keyword}
Time projection chamber \sep Gas electron multiplier \sep High energy heavy ion collision experiment
\PACS 29.40.Cs \sep 29.40.Gx \sep 25.75.-q
\end{keyword}

\end{frontmatter}

\section{Introduction}
High particle multiplicity is an important feature to be considered in detector designs for high energy heavy ion collision experiments. 
In $\sqrt{s_{NN}}=200$~GeV Au+Au collisions at the Relativistic Heavy Ion Collider (RHIC) at Brookhaven National Laboratory, the average charged particle multiplicity $\langle dN_{ch}/d\eta |_{\eta =0}\rangle$ is 170~\cite{cite_1}. 
Therefore, average charged particle density is 0.03~cm$^{-2}$ at a distance of 30~cm from the vertex. 
Additionally, experiments at RHIC are performed at a high event rate of about 10~kHz and the charged particle rate is 300~cps/cm$^2$ at a distance of 30~cm from the vertex. 
This harsh environment demands highly efficient central tracking detectors. 

A wide variety of observables are measured in high energy heavy ion collision experiments: 
Such as charged particle multiplicities, yield ratios and spectra of identified hadrons, elliptic flow, suppression of high $p_T$ particle production and heavy flavor production~\cite{cite_2}. 
Therefore, several particular features are required for the detectors used. 
A relatively wide transverse momentum range ($p_T\sim$~0.2--20~GeV/$c$) is required to be covered to take in the broad interests of high energy heavy ion collisions and the magnetic field should be kept low ($\sim$1~T). 
However, good momentum resolution of $\delta p_T/p_T^2\sim 3 \times 10^{-3}$~(GeV/$c$)$^{-1}$ is required for future measurements, such as the $\Upsilon$ state measurement at RHIC-II~\cite{cite_3}. 
To achieve such good momentum resolution with a magnetic field of 1~T, a 1-m radius solenoidal tracker with spatial resolution of $\sim$200~$\mu$m is needed. 
Double track resolution is also required to be better than 1~cm to cope with the high particle multiplicity. 
Such high flux operational and multi hit capabilities have recently also been required in particle physics experiments~\cite{cite_4}. 

One of the candidates as a detector that satisfies the above requirements is a combination of a time projection chamber (TPC), a sophisticated detector for particle tracking and particle identification, and micro patter gas detectors (MPGDs)~\cite{cite_5} such as gas electron multipliers (GEMs)~\cite{cite_6}. 
Although existing wire chambers need gating wires to collect positive ions and have limitations in their double track resolution by the wire spacing, the novel structure of a GEM has the following advantages in its application to TPC readout:  
\begin{itemize}
\item Two dimensional symmetry of the GEM structure provides a large flexibility in the readout geometry. 
\item Large areas can be covered with a low amount of material since support mechanics for a GEM are simple. 
\item The intense electric field region will be limited inside the holes and the {\bf E}$\times${\bf B} effect is expected to be reduced. (In a strong magnetic field, this effect results in a broadening of the electron cloud and a worsening of the resolution.) 
\item The signal of the readout pad is induced by amplified electrons, and is spatially narrow and fast. 
\item The generated ions do not affect the signal and the ion tail of the signal will be suppressed. 
\item The positive ion feedback into the drift region can be suppressed by a factor of about 10 due to the electric field around the GEM holes and so gating wires might be unnecessary. 
\end{itemize}
Therefore, a TPC using a GEM for signal amplification (GEM-TPC) may achieve high rate capability as well as excellent double track and spatial resolution. 
A GEM-TPC is a strong candidate to be a central tracking detector in high energy heavy ion collision experiments~\cite{cite_7, cite_8, cite_9, cite_10, cite_11}. 

\section{GEM-TPC Prototype}
\subsection{Mechanical Structure}
\begin{figure}[htbp]
  \begin{center}
    \includegraphics[width=13.0cm]{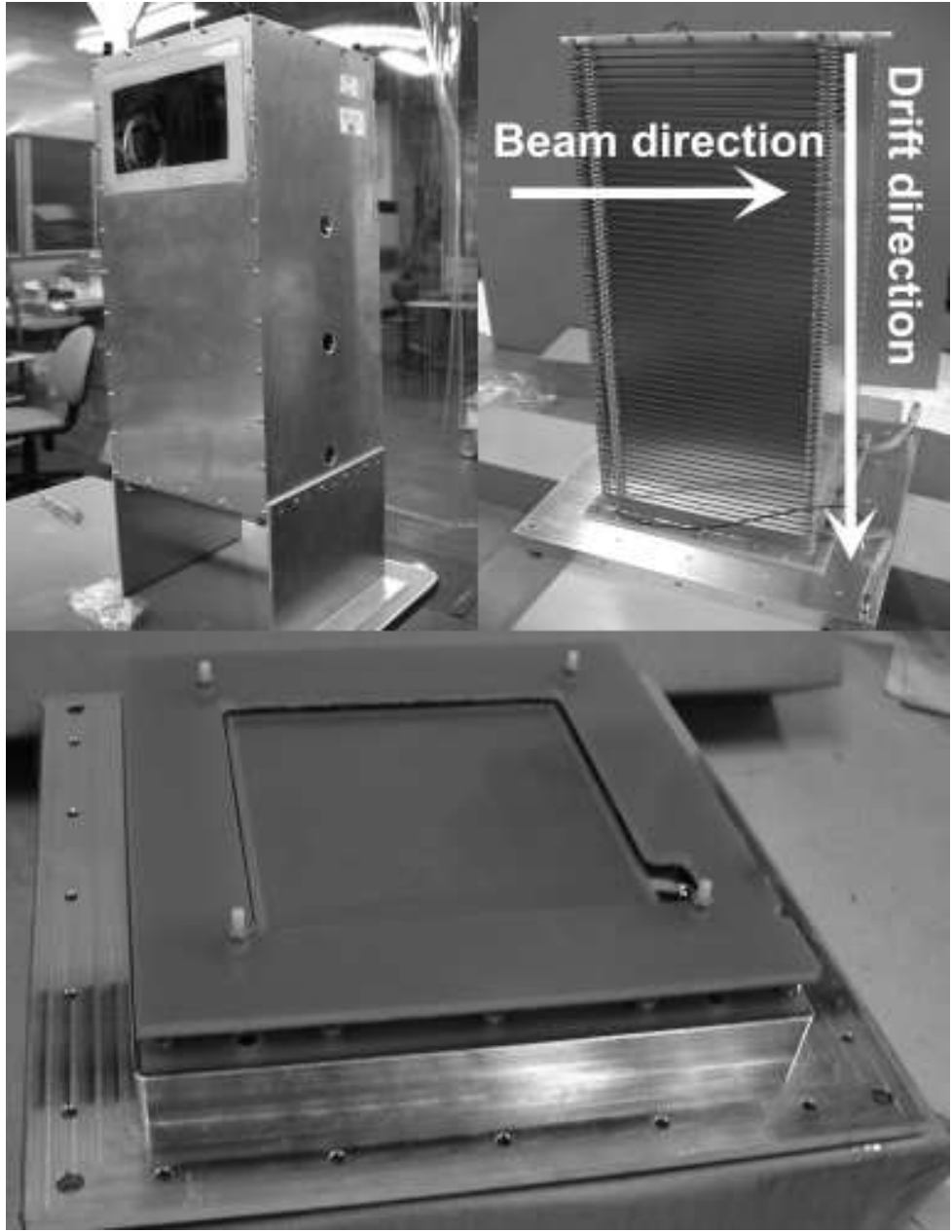}
    \caption{The gas vessel of the GEM-TPC (top left), the field cage (top right) and the end cap chamber (bottom). }
    \label{fig_1}
  \end{center}
\end{figure}

\begin{figure}[htbp]
  \begin{center}
    \includegraphics[width=13.0cm]{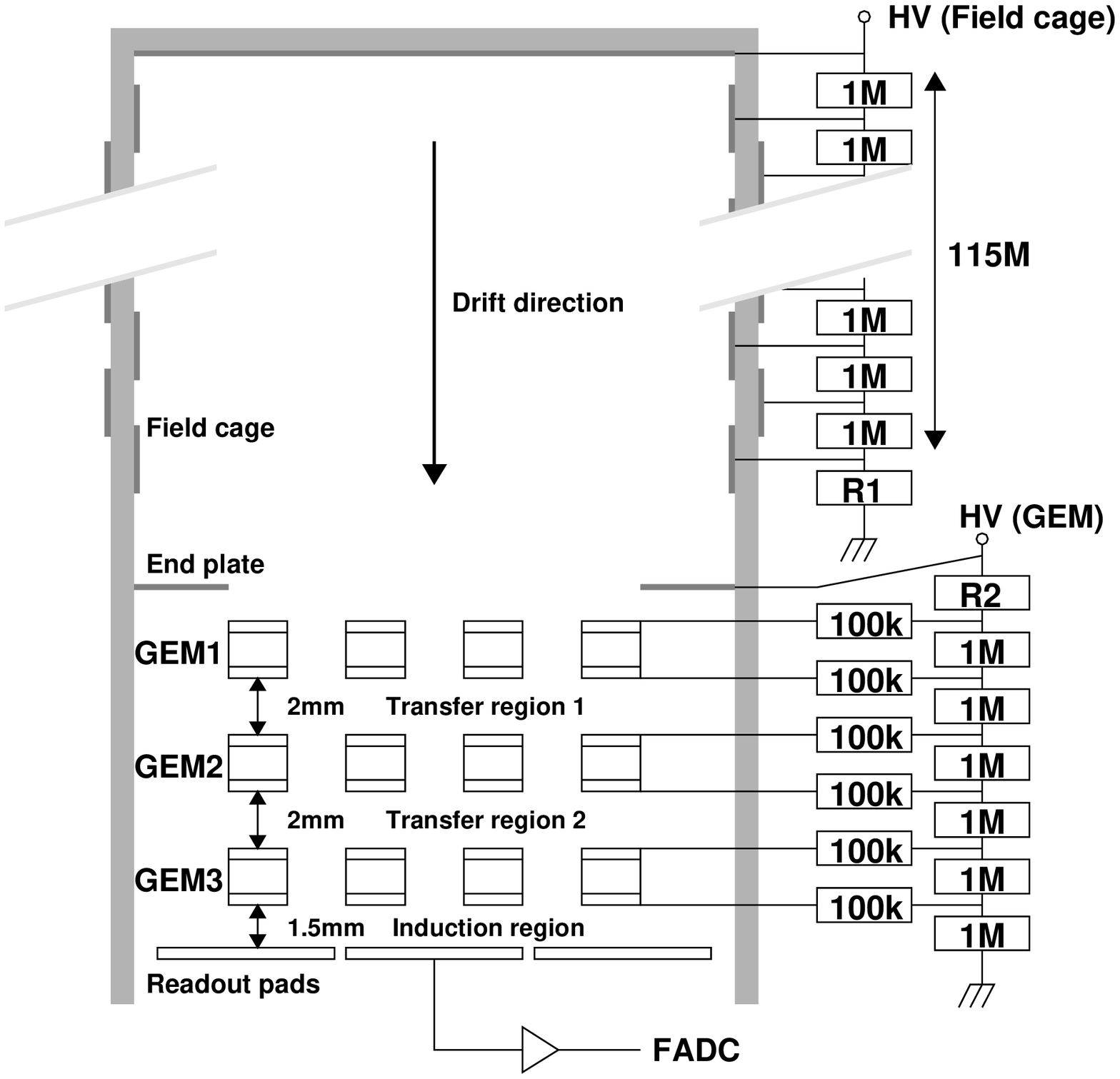}
    \caption{A schematic view of the GEM-TPC. }
    \label{fig_2}
  \end{center}
\end{figure}

A GEM-TPC prototype, consisting of an end cap chamber, a gas vessel and a field cage, was developed~\cite{cite_12}. 
Photographs of the GEM-TPC prototype are shown in Fig.~\ref{fig_1}. 
Figure~\ref{fig_2} shows a schematic view of the GEM-TPC. 

\begin{figure}[htbp]
  \begin{center}
    \includegraphics[width=13.8cm]{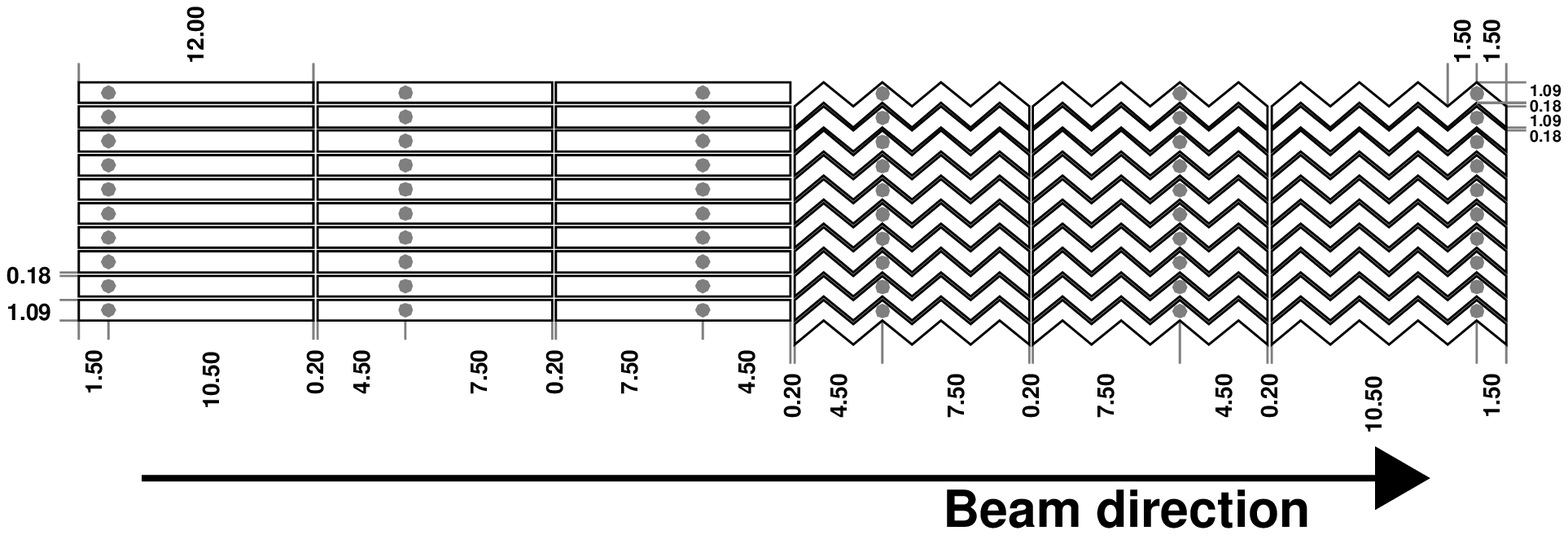}
    \caption{Readout pad layout. There are 3 rows~$\times$~10 columns for both rectangular and chevron pads. The outer 2 columns are not read out. Each pad has an area of 1.09$\times$12.0~mm$^2$. }
    \label{fig_3}
  \end{center}
\end{figure}

The end cap chamber mounts a triple GEM (the effective area is $10 \times 10$~cm$^2$) on readout pads. 
The triple GEM was constructed from GEM foils made at CERN~\cite{cite_13}. 
As shown in Fig.~\ref{fig_2}, the gap between neighboring GEMs was 2~mm and the gap between the bottom GEM and the pads was 1.5~mm. 
High voltages are applied to the triple GEM through connectors penetrating the end cap chamber with a resistor chain. 
When the voltage across the GEMs ($V_{GEM}$) is 360~V, the electric fields in the transfer region and the induction region are $E_t$=360~V/2~mm=1.8~kV/cm and $E_i$=360~V/1.5~mm=2.4~kV/cm, respectively. 
The operated drift electric field is shown in Table~\ref{table_1}. 
Two kinds of readout pads with different shapes, rectangle and chevron (zigzag), were used to study the dependence of the spatial resolution on shape (see Fig.~\ref{fig_3}). 
Since chevron pads may increase the number of hits on pads by charge sharing, chevron pads are expected to have better spatial resolution than rectangular ones~\cite{cite_14}. 
Both kinds of pads, both made of gold-plated copper, have the same area of $1.09 \times 12.0$~mm$^2$ and the same pitch of 1.27~mm. 
Relatively narrow width pads are required for charge sharing for small diffusion gases such as CF$_4$~\cite{cite_12, cite_15}. 

The field cage is a cuboid with dimensions of 36~(drift direction)~$\times$~17~$\times$~17~cm$^3$ and creates a uniform electric drift field. 
The electric field uniformity is \\$\int_{\mbox{{\scriptsize 0~cm}}}^{\mbox{{\scriptsize 30~cm}}}(E_\perp/E)dz\le~1~\mbox{mm}$ in the center area of $10\times 10$~cm$^2$ which corresponds to the GEM effective area. 
The field cage consists of 115 gold-plated copper strips connected with 1-M$\Omega$ resistors in series on FR4 boards. 
At the end of the resistor chain, additional resistors are placed to match the voltage at the bottom of the field cage with the surface voltage of the top GEM. 

The gas vessel is made of aluminum plates and has a volume of $60 \times 29 \times 29$~cm$^3$. 

\subsection{Front End Electronics}
A charge sensitive preamplifier, consisting of two kinds of operational amplifiers (AD8058 and AD8132, Analog Devices, Inc.), was used for the GEM-TPC. 
Its time constant is $\tau=1~\mu$s and its effective gain $G=3.2~\mbox{V}/\mbox{pC}$. 
The gain of the preamplifier was determined from the expected signal amplitude and the dynamic range ($0\sim -1$~V) of a flash ADC (FADC) module (RPV-160, REPIC Co., Ltd.). 
The resolution and the sampling rate of the FADC are 8~bits and 100~MHz, respectively. 
Signals from 24 pads (3 rows~$\times$~8 columns) are transmitted from the preamplifiers to the FADCs through 8-m shielded twisted cables. 

\subsection{Gas}
Three kinds of gases with different properties, a mixture of argon(Ar)(90\%)-methane(CH$_4$)(10\%) (commonly called P10), a mixture of argon(Ar)(70\%)-ethane(C$_2$H$_6$)(30\%) and pure tetrafluoromethane(CF$_4$)(99.999\%) gas, were used to study the performance of the GEM-TPC. 
Properties of these gases are shown in Table~\ref{table_1}. 
Drift velocities and diffusion coefficients were calculated by Magboltz~\cite{cite_16}. 

\begin{table}
  \caption{Properties of the 3 kinds of gases. Drift velocities and diffusion coefficients were calculated by Magboltz. }
  \label{table_1}
  \begin{center}
    \begin{tabular}{cccc}
      \hline
      Gas & Ar(90\%)-CH$_4$(10\%) & Ar(70\%)-C$_2$H$_6$(30\%) & CF$_4$\\
      \hline
      Operated drift field (V/cm) & 130 & 390 & 570\\
      Drift velocity (cm/$\mu$s)& 5.48 & 5.01 & 8.90\\ 
      Transverse diffusion ($\mu$m/$\sqrt{\mbox{cm}}$) & 570 & 306 & 104\\
      Longitudinal diffusion ($\mu$m/$\sqrt{\mbox{cm}}$) & 378 & 195 & 82\\
      Mean energy for ion-electron & 26 & 26 & 54 \\ 
      pair production (eV) & & & \\
      \hline
    \end{tabular}
  \end{center}
\end{table}

Ar(90\%)-CH$_4$(10\%) is widely used in TPCs and GEMs~\cite{cite_11, cite_17, cite_18}. 
A fast drift velocity that peaks at a low electric field of 130~V/cm is the primary attribute of Ar-CH$_4$, and such a low electric field is an advantage for detector operation. 
However, its large diffusion coefficients are a disadvantage. 
Ar(70\%)-C$_2$H$_6$(30\%) is a common chamber gas, but there are few results for it with a GEM. 
Although its diffusion coefficients are smaller than those of Ar-CH$_4$, the drift velocity peaks at a relatively high electric field of 550~V/cm. 
CF$_4$ gas is studied as a TPC and GEM gas because of its very small diffusion coefficients and very fast drift velocity~\cite{cite_12, cite_15}. 
These properties of CF$_4$ will be advantages in high particle multiplicity environments of high energy heavy ion collision experiments. 
However, a high electric field is needed to achieve a fast drift velocity with CF$_4$. 

The gas flow rate was set at $\sim$200~ml/min using a mass flow controller (SEC-E40, ESTEC Co., Ltd.). 
Gas pressure was set at the atmospheric pressure using a bubbler filled with silicone oil. 

\begin{figure}[htbp]
  \begin{center}
    \includegraphics[width=13.0cm]{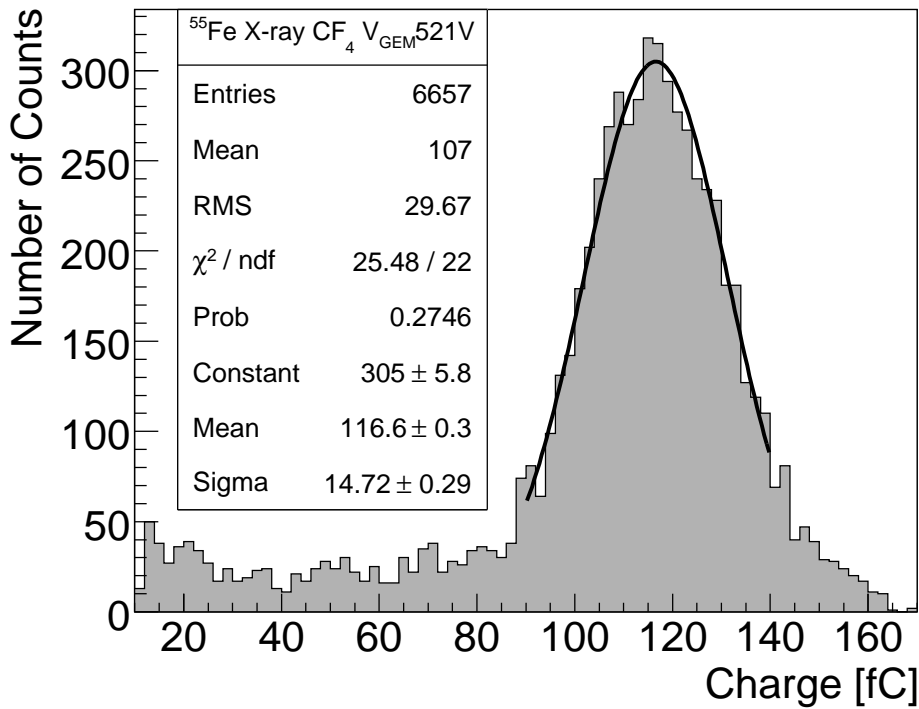}
    \caption{Measured $^{55}$Fe X-ray (5.9~keV) charge spectrum with CF$_4$. The energy resolution is $\sigma_E/E=13\%$. }
    \label{fig_4}
  \end{center}
\end{figure}

\begin{figure}[htbp]
  \begin{center}
    \includegraphics[width=13.0cm]{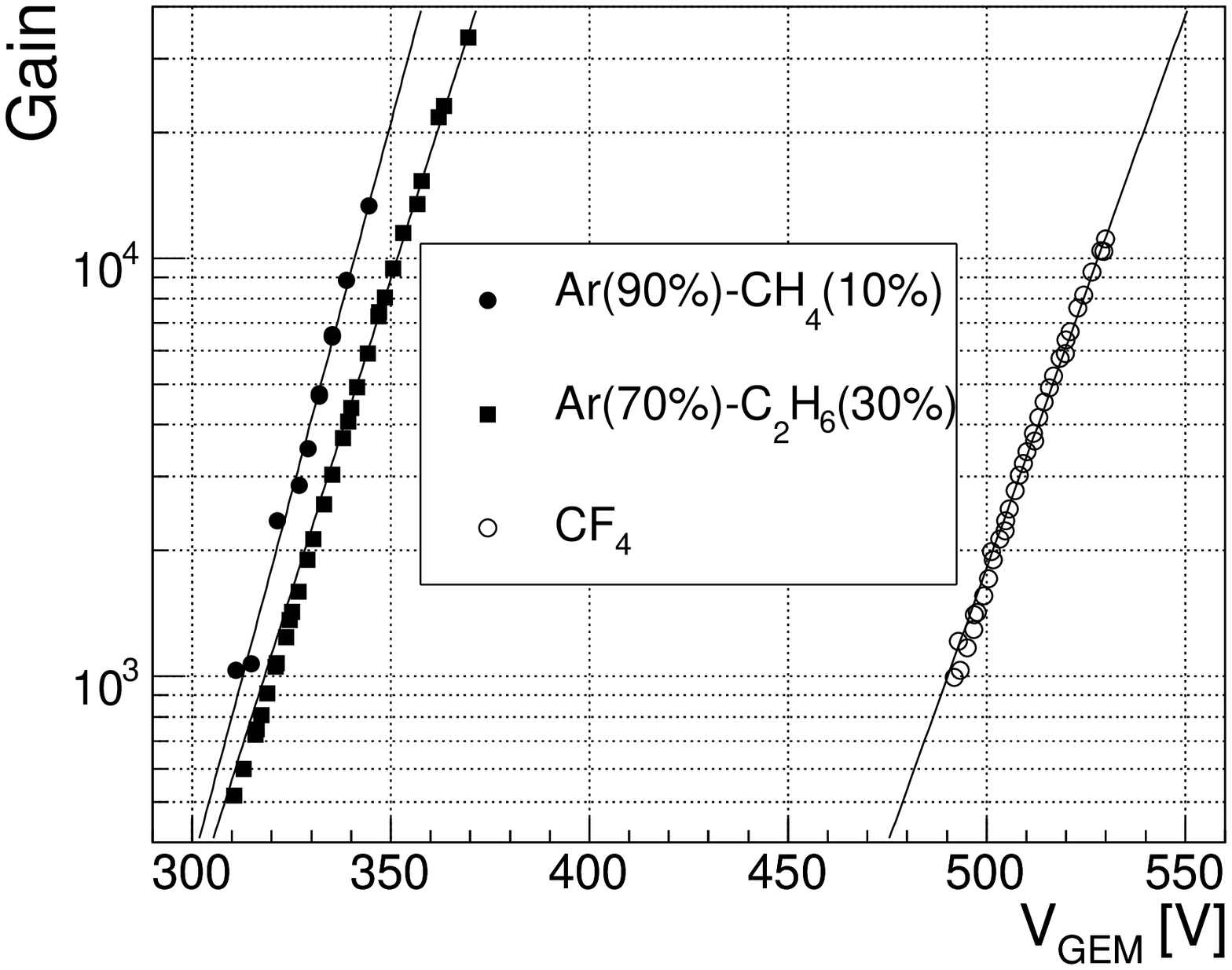}
    \caption{Measured gain curves with the 3 kinds of gases (Ar(90\%)-CH$_4$(10\%), Ar(70\%)-C$_2$H$_6$(30\%) and CF$_4$) as functions of the voltage across the GEMs, $V_{GEM}$. }
    \label{fig_5}
  \end{center}
\end{figure}

Effective gas gains were measured using an $^{55}$Fe X-ray (5.9~keV) source. 
An X ray creates primary electrons by a photoelectric effect with a gas molecule. 
The obtained $^{55}$Fe X-ray charge spectrum with CF$_4$ is shown in Fig.~\ref{fig_4}. 
The spectrum around the 5.9-keV peak is fitted by a Gaussian function and the gain $G$ is calculated as 
\begin{eqnarray}
G=\frac{\langle Q\rangle}{e}\cdot\frac{W}{5.9~\mbox{keV}},
\end{eqnarray}
where $\langle Q\rangle$ stands for the mean of the Gaussian function, $e$ is the electron charge, and $W$ is the mean energy for ion-electron pair production and is given in Table~\ref{table_1}. 
The gains for the 3 kinds of gases are shown as a function of $V_{GEM}$ in Fig.~\ref{fig_5} and a gain of $10^4$ was achieved with each gas. 
Although CF$_4$ needs a higher voltage than Ar-CH$_4$ and Ar-C$_2$H$_6$ to achieve the same gain, the slopes of the gain curves are similar for the 3 gases. 
From the obtained Gaussian sigma, $\sigma_E$, the relative energy resolution ($\sigma_E/E$) with Ar-CH$_4$, Ar-C$_2$H$_6$ and CF$_4$ is determined to be 11\%, 10\% and 13\%, respectively. 
The obtained energy resolution is comparable with the result by another research group (16\% for CF$_4$ and 9\% for Ar(70\%)-CO$_2$(30\%))~\cite{cite_19}. 

\section{Performance Test}
A beam test was performed at the $\pi$2 secondary beam line of the 12-GeV Proton Synchrotron at KEK (KEK-PS) to evaluate the basic performance of the GEM-TPC. 

The characteristics of the GEM-TPC evaluated in the performance test were detection efficiency, spatial resolution in the pad-row direction and the drift direction and particle identification capability by $dE/dx$ measurement. 
The dependence of these characteristics on the 3 kinds of gases (Ar(90\%)-CH$_4$(10\%), Ar(70\%)-C$_2$H$_6$(30\%) and CF$_4$), the GEM gain ($6\times 10^2$--$2\times 10^4$), the drift length (20--290~mm), the readout pad shape (rectangle and chevron), the beam momentum (0.5--3.0~GeV/$c$) and the beam rate was also evaluated. 
Except for the GEM gain dependence measurement, $V_{GEM}$ was fixed to 332~V, 341~V and 498~V for Ar-CH$_4$, Ar-C$_2$H$_6$ and CF$_4$, respectively. 

\begin{figure}[htbp]
  \begin{center}
    \includegraphics[width=14.2cm]{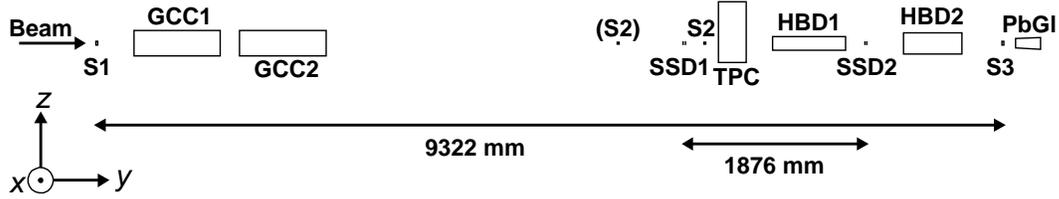}
    \caption{A schematic view of the detector setup in the performance test. }
    \label{fig_6}
  \end{center}
\end{figure}

Figure~\ref{fig_6} shows a schematic view of the detector setup for the performance test. 
Three plastic scintillation counters (S1, S2 and S3) were used for event triggering and time of flight measurements for particle identification. 
Two gas Cherenkov counters (GCC1 and GCC2) filled with 2.5-atm CO$_2$ gas were used for electron identification. 
Two silicon strips detectors (SSD1 and SSD2), each of which consists of two single sided strips (strip pitch of 80~$\mu$m) crossing at right angles, were used for particle tracking. 
Two hadron blind detectors (HBD1 and HBD2) were tested at the same time and the results are shown in~\cite{cite_20}. 
The GEM-TPC was operated without a magnetic field. 

\begin{figure}[htbp]
  \begin{center}
    \includegraphics[width=13.0cm]{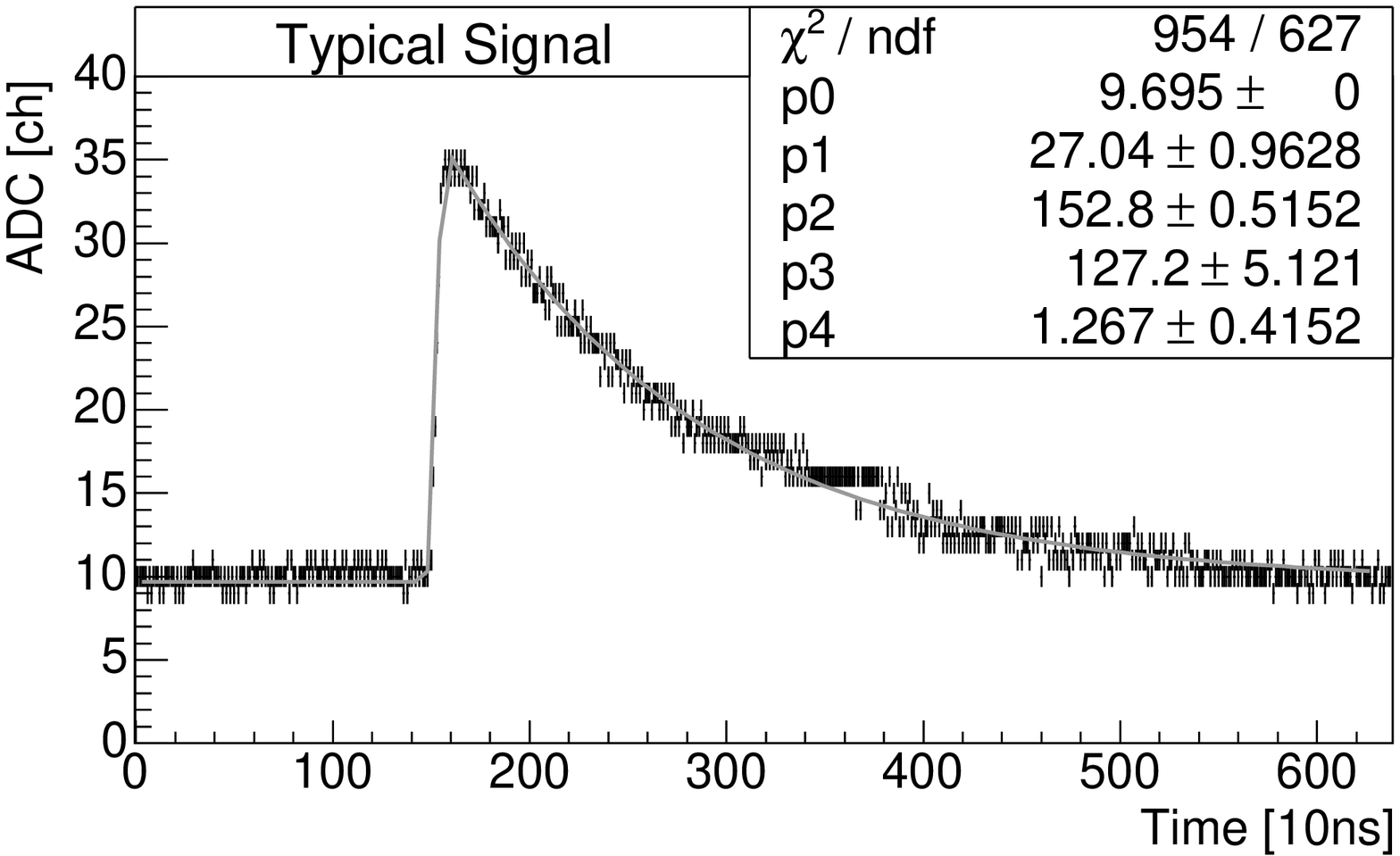}
    \caption{A typical GEM-TPC signal crated by a 1-GeV/$c$ electron beam. The drift gas is Ar(70\%)-C$_2$H$_6$(30\%), the drift length is 85~mm and the pad shape is chevron. A voltage across the GEMs is $V_{GEM}=341$~V. } 
    \label{fig_7}
  \end{center}
\end{figure}

Figure~\ref{fig_7} shows a typical signal of the GEM-TPC operated with Ar-C$_2$H$_6$. 
The signal was recorded for 6.4~$\mu$s~(=640~samples) in one event. 
One channel of the ADC corresponds to $\sim~4$~mV. 
To extract the pulse height and the arrival time of the signal, the following function is fitted to the FADC spectrum, 
\begin{eqnarray}
ADC(t) = p_0+\frac{p_1\cdot\exp(-(t-p_2)/p_3)}{1+\exp(-(t-p_2)/p_4)},
\end{eqnarray}
where $t$ is sampling time. 
The fitting parameters in the above function can be recognized as follows: $p_0$ is the pedestal, $p_1$ is the pulse height, $p_2$ is the arrival time, $p_3$ is the time constant of the electronics and $p_4$ is the rise time. 
The obtained pulse height and arrival time are used for determinations of the hit position. 
The hit position in each pad row is determined by an amplitude weighted mean of pad positions and arrival times: 
\begin{eqnarray}
x_i=\frac{\sum_{j=0}^{8-1}p_{1,i,j}\cdot j \cdot D}{\sum_{j=0}^7p_{1,i,j}}
\end{eqnarray} 
in the pad-row direction and 
\begin{eqnarray}
z_i=\frac{\sum_{j=0}^{8-1}p_{1,i,j}\cdot p_{2,i,j} \cdot v_{drift}}{\sum_{j=0}^7p_{1,i,j}}
\end{eqnarray} 
in the drift direction, where $p_{k,i,j}$ is the $k$-th parameter of the $i$-th row and the $j$-th column pad ($0\le i <3$ and $0\le j <8$), $D=1.27$~mm is the pad spacing (see Fig.~\ref{fig_3}) and $v_{drift}$ is the drift velocity.   

\section{Results}
\subsection{Detection Efficiency}
\begin{figure}[htbp]
  \begin{center}
    \includegraphics[width=13.0cm]{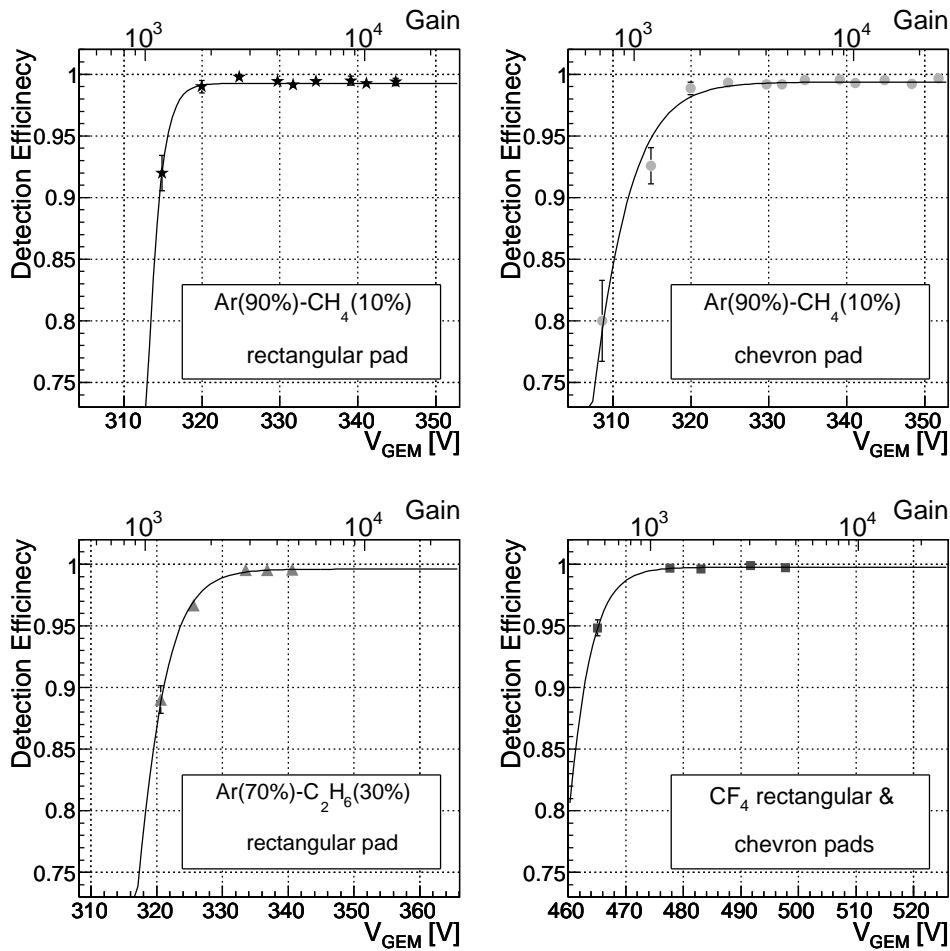}
    \caption{Detection efficiency of the GEM-TPC as a function of a GEM gain. }
    \label{fig_8}
  \end{center}
\end{figure}

Single-pad-row detection efficiency was measured as a function of $V_{GEM}$ with the 3 kinds of gases. 
Measurements were done with 1-GeV/$c$ $\pi^-$ beams with a drift length of 20~mm with Ar-CH$_4$ and 85~mm with both Ar-C$_2$H$_6$ and CF$_4$. 
Tracks having hits in the 1st and 3rd pad rows were selected for the efficiency evaluation, and the fraction of the hits in the 2nd pad row was used as the detection efficiency. 
Results are shown in Fig.~\ref{fig_8}. 
The detection efficiency reaches a plateau at a gain of $\sim 4 \times 10^3$ with Ar-CH$_4$ and Ar-C$_2$H$_6$. 
The small diffusion of CF$_4$ makes the efficiency reach a plateau at a smaller gain of $\sim 2 \times 10^3$.  
The efficiency plateaus are 99.3\%, 99.6\% and 99.8\% with Ar-CH$_4$, Ar-C$_2$H$_6$ and CF$_4$, respectively. 
These results are very similar to results from another research group findings of Ar(70\%)-CO$_2$(30\%) and Ar(93\%)-CH$_4$(5\%)-CO$_2$(2\%)~\cite{cite_10}. 

The gain scales for Ar-CH$_4$ and Ar-C$_2$H$_6$ in Fig.~\ref{fig_8} are determined by the obtained gain-$V_{GEM}$ relation in Fig.~\ref{fig_5} and the one for CF$_4$ is determined by measured pulse heights using the relation between the gains and measured pulse heights with Ar-CH$_4$ and Ar-C$_2$H$_6$. 
The gain with CF$_4$ in Fig.~\ref{fig_8} is larger than the one in Fig.~\ref{fig_5}. 
This difference may be due to differences of the atmospheric pressure $p$ and the temperature $T$ because the GEM gain strongly depends on $p/T$~\cite{cite_19}. 
Unfortunately, the atmospheric pressure and the temperature were not monitored during the performance test. 

\subsection{Transverse Diffusion Coefficient}
\begin{figure}[htbp]
  \begin{center}
    \includegraphics[width=13.0cm]{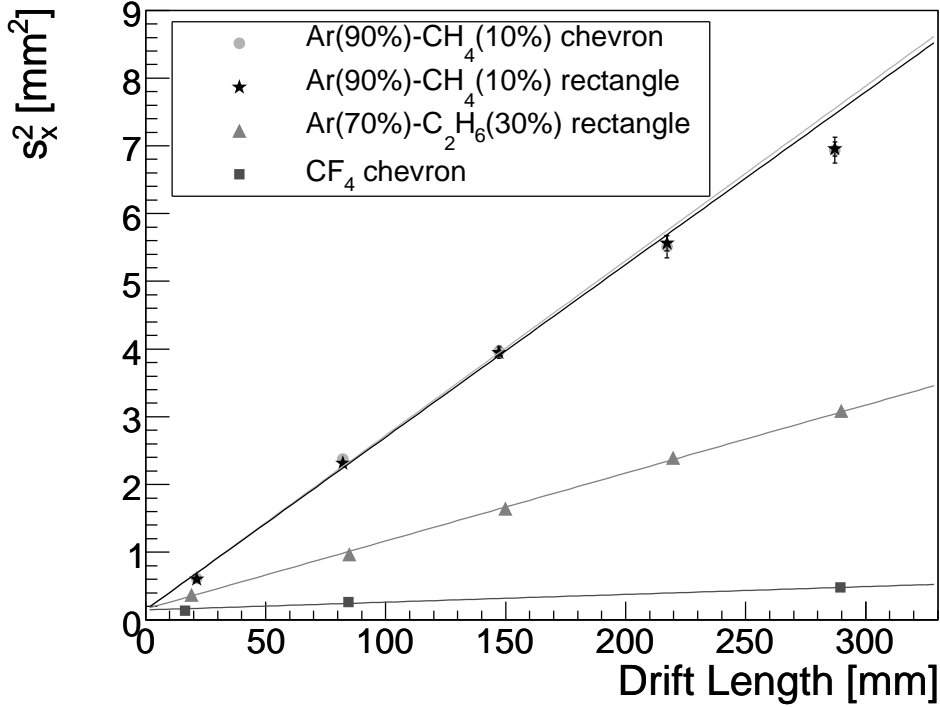}
    \caption{The squared width of the diffused secondary electrons $s_x^2$ for the 3 kinds of gases. }
    \label{fig_9}
  \end{center}
\end{figure}

Transverse diffusion coefficients were measured with 1-GeV/$c$ $\pi^-$ beams. 
The coefficients are evaluated using the spatial distribution of secondary electrons in the pad-row direction. 
The secondary electron distribution is fitted by a Gaussian distribution. 
The obtained sigma of Gaussian, $s_x$, can be expressed as 
\begin{eqnarray}
s_x^2(L)=s_{x0}^2+C_{DT}^2\cdot L,
\end{eqnarray}
where $L$ is the drift length, $C_{DT}$ is the transverse diffusion coefficient and $s_{x0}$ is the intrinsic width of the induced charge distribution determined by the readout system configuration. 
Figure~\ref{fig_9} shows $s_x^2$ as a function of the drift length. 
The measured $s_{x0}$ and $C_{DT}$ are shown in Table~\ref{table_2}. 
The measured values of $C_{DT}$ of Ar-C$_2$H$_6$ and CF$_4$ agree well with the calculated values of $C_{DT}$ shown in Table~\ref{table_1}; however, for Ar-CH$_4$ the measured one is slightly smaller than the calculated one. 

\begin{table}
  \caption{The measured width of the induced charge distribution and transverse coefficients. Only statistical errors are shown. } 
  \label{table_2}
  \begin{center}
    \begin{tabular}{cccc}
      \hline
      Gas & Pad shape & $s_{x0}$ ($\mu$m)  & $C_{DT}$ ($\mu$m/$\sqrt{\mbox{cm}}$) \\
      \hline
      Ar(90\%)-CH$_4$(10\%) & chevron & $385~\pm~101$ & $508~\pm~7$ \\
      Ar(90\%)-CH$_4$(10\%) & rectangle & $387~\pm~101$ & $505~\pm~7$ \\
      Ar(70\%)-C$_2$H$_6$(30\%) & rectangle & $402~\pm~43$ & $317~\pm~4$ \\
      CF$_4$ & chevron   & $383~\pm~37$ & $107~\pm~6$ \\
      \hline
    \end{tabular}
  \end{center}
\end{table}

\subsection{Spatial Resolution}
\begin{figure}[htbp]
  \begin{center}
    \includegraphics[width=13.0cm]{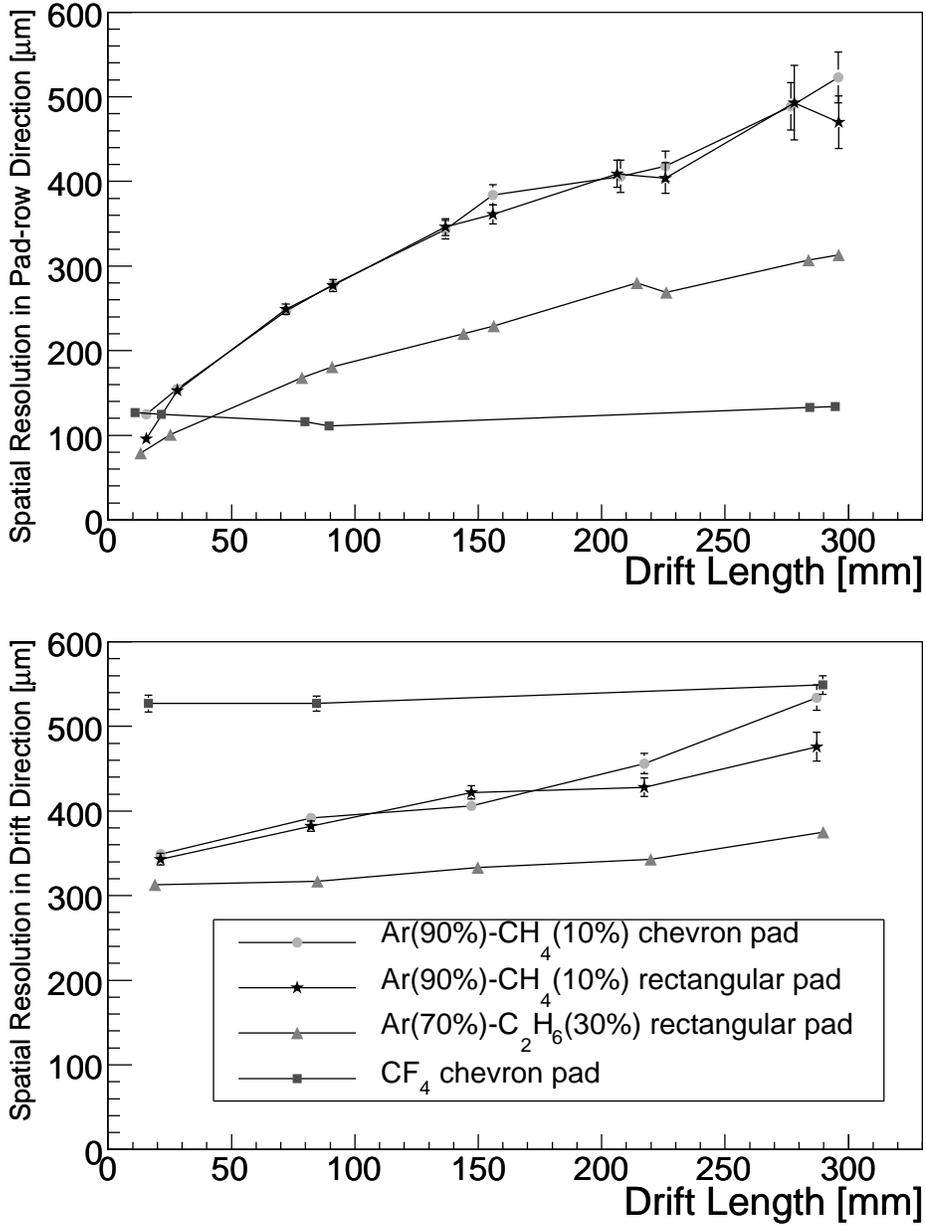}
    \caption{Spatial resolution in the pad-row direction (top) and the drift direction (bottom). }
    \label{fig_10}
  \end{center}
\end{figure}

Single-pad-row spatial resolution in the pad-row and drift directions was evaluated for a drift length range of 20--290~mm with 1-GeV/$c$ $\pi^-$ beams. 
The single-pad-row spatial resolution in both directions was evaluated by the residual between the hit position of the 2nd pad row and the interpolated hit position from the 1st and 3rd pad rows. 
The measured spatial resolution is shown in Fig.~\ref{fig_10}. 
The effect of diffusion on the spatial resolution in both directions is clearly seen. 
The best resolution is 79~$\mu$m in the pad-row direction and 313~$\mu$m in the drift direction, obtained with Ar-C$_2$H$_6$ gas and rectangular pads at 13-mm drift. 
The spatial resolution of the chevron pads is almost the same as that of the rectangular ones. 
If the charge distribution is Gaussian, the spatial resolution of the chevron pads should be better than that of the rectangular ones~\cite{cite_14}.
A possible reason for the result obtained is that the finite sizes of the GEM holes distort the charge distribution from a Gaussian distribution and so the non-Gaussian tails worsen the spatial resolution of the chevron pads. 

The dependence of the spatial resolution in the pad-row direction on the drift length of $L$ can naively be understood as 
\begin{eqnarray}
\sigma_x^2(L)\ =\ \sigma_{x0}^2+C_{DT}^2\cdot L/N_{eff},
\end{eqnarray}
where $\sigma_{x0}$ is the extrapolated resolution at a zero drift length and $N_{eff}$ is the effective number of secondary electrons~\cite{cite_11}. 
$N_{eff}$ is mainly determined by the number of secondary electrons, $N$, and pad geometry~\cite{cite_21}. 
The calculated number of secondary electrons, $N$, and the measured number of effective secondary electrons, $N_{eff}$, for 1-GeV/$c$ $\pi^-$ beams and a track length of 12~mm are shown in Table~\ref{table_3}. 
A small fraction of the secondary electrons effectively contribute to spatial resolution. 

\begin{table}
  \caption{The number of effective secondary electrons, $N_{eff}$, and the number of secondary electrons, $N$, for 1-GeV/$c$ $\pi^-$ beams and the track length of 12~mm. }
  \label{table_3}
  \begin{center}
    \begin{tabular}{ccccc}
      \hline
      Gas & Pad shape & $N_{eff}$  & $N$ & $N_{eff}/N$           \\
      \hline
      Ar(90\%)-CH$_4$(10\%) & chevron   & 31$\pm$ 1 & 119 & 0.26$\pm$0.01\\
      Ar(90\%)-CH$_4$(10\%) & rectangle & 30$\pm$ 1 & 119 & 0.25$\pm$0.01\\ 
      Ar(70\%)-C$_2$H$_6$(30\%) & rectangle & 31$\pm$ 1 & 131 & 0.23$\pm$0.01\\ 
      CF$_4$ & chevron & 86$\pm$23 & 147 & 0.58$\pm$0.15\\
      \hline
    \end{tabular}
  \end{center}
\end{table}

\subsection{Beam Rate Dependence}
\begin{figure}[htbp]
  \begin{center}
    \includegraphics[width=13.0cm]{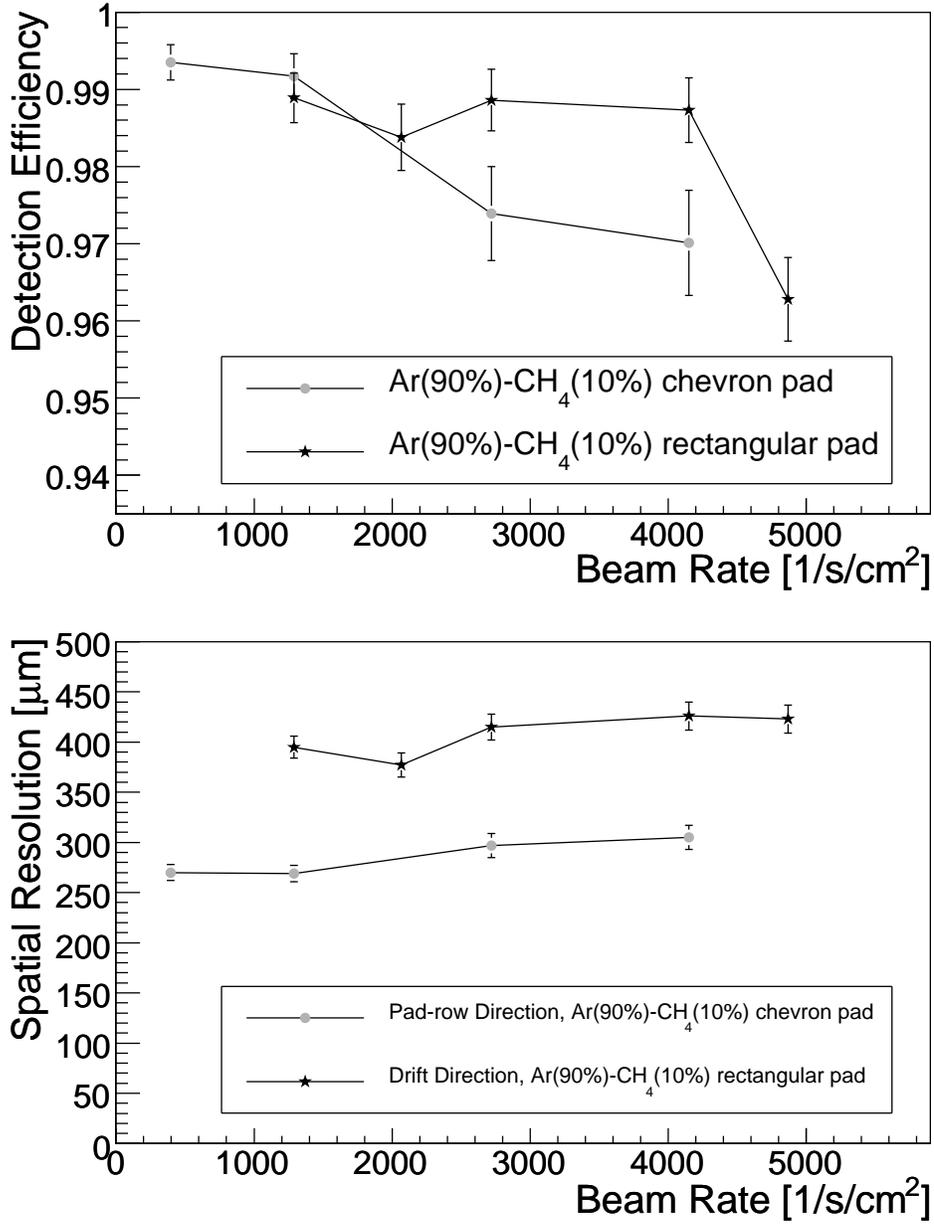}
    \caption{Dependence of the detection efficiency (top) and the spatial resolution on the beam rate. }
    \label{fig_11}
  \end{center}
\end{figure}

One of the advantages of the GEM-TPC is its ion feedback suppression. 
The effect of ion feedback on GEM-TPC performance was studied by measuring the beam rate dependence of the detection efficiency and spatial resolution. 
The beam rate was determined by the beam slit width and the rate was monitored with the S2 scintillator (2.5$\times$2.5~cm$^2$). 
The beams of $e^+$, $\pi^+$ and $p$ at a momentum of 2~GeV/$c$ and Ar-CH$_4$ gas were used, while the drift length was 85~mm.  
The results are shown in Fig.~\ref{fig_11}. 
The results of the detection efficiency and the spatial resolution in the previous subsections were obtained with a beam rate of $\le 500$~cps/cm$^2$.  
At the maximum beam rate of 4800~cps/cm$^2$, the detection efficiency and the spatial resolution were worsened by factors of 2.5$\pm$0.5\% and 11$\pm$3\%, respectively. 
The maximum total beam rate in the active GEM-TPC region was in the order of 10$^5$~cps. 

This result is worse than the results from the other research group with beams of 1.5$\times 10^{5}$~cps/cm$^2$~\cite{cite_7}. 
Because of the limited readout area of our GEM-TPC (1$\times$3.6~cm$^2$), it was not possible to fully distinguish double tracks, which worsened the detection efficiency and the spatial resolution. 

The beam rate exceeds the typical rates in $\sqrt{s_{NN}}$=200~GeV Au+Au collisions at RHIC and $\sqrt{s_{NN}}$=5.5~TeV Pb+Pb collisions at LHC, which are 300~cps/cm$^2$ and 1400~cps/cm$^2$, respectively, at a distance of 30~cm from the vertex. 
Since 4800~cps/cm$^2$ is much larger than these numbers, the effect of the ion feedback on the GEM-TPC performance can be regarded as negligible for our purpose. 

\subsection{Particle Identification Using $dE/dx$}
\begin{figure}[htbp]
  \begin{center}
    \includegraphics[width=13.0cm]{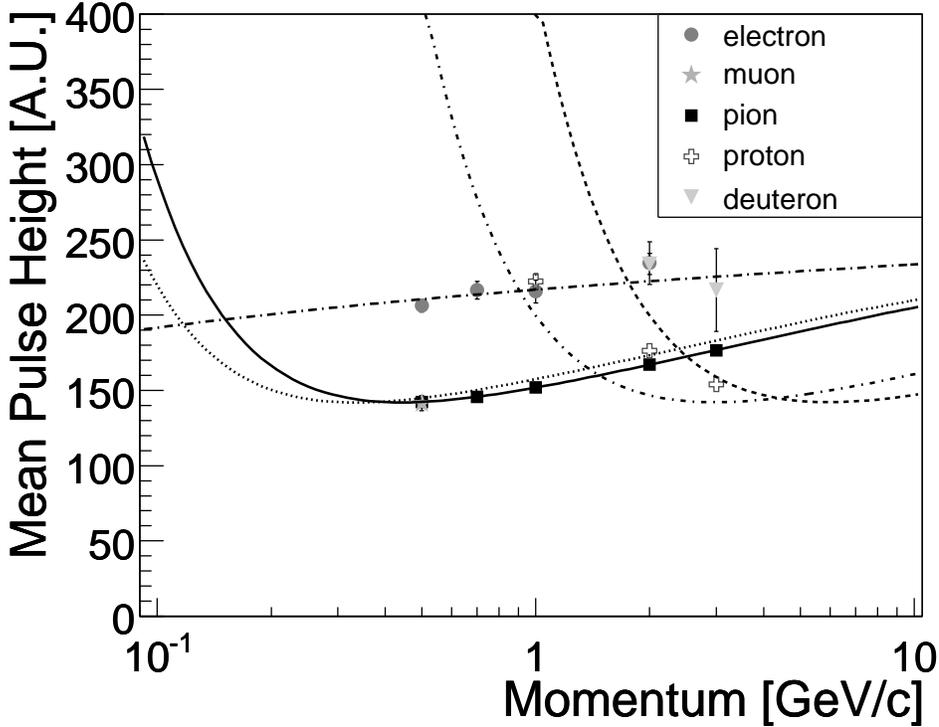}
    \caption{Corrected mean pulse heights for 5 kinds of particle species. The curves are the expected mean pulse heights. }
    \label{fig_12}
  \end{center}
\end{figure}

Energy losses, $dE/dx$, were measured for positrons, muons, pions, protons and deuterons in a beam momentum range of 0.5--3.0~GeV/$c$ to evaluate the particle identification capability. 
In this measurement, the drift length was 85~mm and Ar-CH$_4$ was used. 
The summation of pulse heights for 3 pad rows is regarded as the energy loss. 
Since there was a time variation in the GEM gain (30\% peak to peak, for 5 hours), the measured value of the mean pulse height, $\langle PH\rangle_{meas}^{i}$ ($i =e^+, \mu^+, \pi^+, p, d$), was corrected with pions at each momentum to eliminate the effect of this time variation as follows. 
\begin{eqnarray}
\langle PH\rangle_{corr}^{i}=\langle PH\rangle_{meas}^{i}\cdot\frac{\langle PH\rangle_{exp}^{\pi^+}}{\langle PH\rangle_{meas}^{\pi^+}}, 
\end{eqnarray}
where $\langle PH\rangle_{corr}^i$ is the corrected mean pulse height and $\langle PH\rangle_{exp}^i$ is the expected mean pulse height. 
Figure~\ref{fig_12} shows the corrected mean pulse heights and curves of the expected mean pulse heights. 
As previously mentioned, the gain variation may be due to the change of $p/T$. 
Another candidate as a reason for the gain variation is the charging-up of the insulator of a GEM foil~\cite{cite_22}. 

To estimate the particle identification capability of a large GEM-TPC, a Monte-Carlo simulation was performed using measured pulse height spectra for 1-GeV/$c$ pions and protons. 
To improve the energy resolution, a truncated mean method, where 2/7 of the pad rows having the largest signals are removed, was used. 
Energy resolution of pions will be 9\% and the pion rejection factor with 99\% proton efficiency is expected to be 200 with a 50-cm track length. 
This energy resolution is comparable with that of the STAR TPC (8\%) with a track length of more than 67~cm~\cite{cite_17}. 

\section{Conclusion}
A GEM-TPC prototype was constructed to develop a tracking detector for use in high event rate and high particle multiplicity environments in high energy heavy ion collisions. 

To evaluate the performance of the GEM-TPC, a beam test was performed at KEK. 
Detection efficiency of $\ge$99.3\% was achieved with 3 kinds of gases, Ar(90\%)-CH$_4$(10\%), Ar(70\%)-C$_2$H$_6$(30\%) and CF$_4$. 
Spatial resolution of 79~$\mu$m in the pad row direction and 313~$\mu$m in the drift direction was achieved with Ar-C$_2$H$_6$ and rectangular pads for 13-mm drift. 
The GEM-TPC showed high detection efficiency and good spatial resolution with a particle rate of 4800~cps/cm$^2$, which exceeds the particle rate of RHIC and the LHC. 
Energy loss measurements showed a good particle identification capability. 

These results indicate that the GEM-TPC meets the requirements for central tracking detectors for use in the next generation of high energy heavy ion collision experiments. 

\section{Acknowledgements}
The authors would like to thank the KEK-PS group, especially Dr.~M.~Ieiri, for their good servicing of the accelerator and excellent cooperation. 
The authors are also thankful to the staff and the students of the Weizmann Institute of Science, University of Tsukuba and Hiroshima University for their cooperation in the performance test. 
The authors acknowledge support from Ministry of Education, Culture, Sports, Science, and Technology and the Japan Society for the Promotion of Science.

\end{document}